\documentstyle[aps,pre,preprint,graphics,epsfig]{revtex}

\newcommand{\ber}{\hbox{ber}}
\newcommand{\bei}{\hbox{bei}}

\begin{document}
\draft

\title{Analytic approximations to Kelvin functions 
        with applications to electromagnetics}

\author{L. Brualla\thanks{\tt Llorenc.Brualla@upc.es}}

\address{Departament de F\'{\i}sica i Enginyeria Nuclear, 
        Universitat Polit{\`e}cnica de Catalunya, \\ 
        Sor Eul{\`a}lia d'Anzizu s/n, E--08034 Barcelona, Spain.}

\author{P. Martin}  
\address{Departamento de F\'{\i}sica, Universidad Sim{\'o}n Bol\'{\i}var, 
        Apartado 89000, Caracas 1080A, Venezuela.}
        
\maketitle

\begin{abstract}
We present analytical approximations for the real
Kelvin function $\ber \, x$ and the imaginary Kelvin function $\bei \,
x$, using the two--point quasifractional approximation
procedure. We have applied these approximations to the calculation
of the current distribution within a cylindrical conductor.  Our
approximations are simple and accurate. The infinite number
of roots is also obtained with the approximation and the precision
increases with the value of the root. 
Our results could find useful applications in problems where analytical
approximations of the Kelvin functions are needed.
\end{abstract}

\section {Introduction}
Within the scope of special functions, Kelvin functions appear early
 in the literature \cite{ryzhik} \cite{abram}. They are derived from
 the Bessel functions \cite{gross} of complex argument and many formulas
 can be found. In addition to their importance in Mathematics they have,
 also, a broad spectrum of application in Physics. For instance, in the
 analysis of the current distribution in cylindrical conductors due to
 wave propagation \cite{marion}. A more interesting application is the
 problem of determining the equivalent impedance of a cylindrical
 conductor which is given in terms of $\ber \, x$ and $\bei \, x$.
 Kelvin functions also arose in the problem of the wave--impedance of
 cylindrically layered conductors for application in nondestructive
 testing \cite{wait}. The temperature distribution in cylindrical
 conductors due to an alternative current also involves  $\ber \, x$
 and $\bei \, x$ \cite{rolicz}. Kelvin functions also appear in other
 fields (i.e., fluid mechanics) but in this paper we will concentrate
 on the first electromagnetic application mentioned above.

One of the problems we have found in the applications of Kelvin
 functions is that their behaviour is rather complicated. Kelvin
 functions are oscillatory and the amplitude of the oscillations
 increases rapidly. In spite of the already existing tables, still
 the functions are part of integrands where the tables are difficult
 to use. The programs to compute these functions are also too elaborated.

It seems convenient to have a good approximation to the functions,
 but the conventional approximation methods do not give good results
 because  of the pathological behaviour of the functions. However,
 in this paper it is shown how those difficulties can be surmounted
 by using the recently published method of two--point quasifractional
 approximations \cite{martin} \cite{donoso}. This has allowed us to
 find analytic approximations to Kelvin functions with good accuracy,
 and furthermore, the same analytic approximation is valid for the
 whole range of positive values of the variable.

Power series and asymptotic expansions are required in order to obtain
 these approximations, thus we start analysing those expansions to
 proceed further away. This will be done in Sec. 2, where the form of
 the approximation will be also derived. Later, the material of this
 paper will be arranged as follows: the calculation of the parameters
 for the two--point quasifractional approximation of \( \ber \, x \)
 is carried out in Sec. 3. Sec. 4 devotes its attention to the analysis
 of the results of the approximation. The same procedure applied to
 \( \ber \, x \) in previous sections is employed in the calculation
 of \( \bei \, x \). That calculation and the discussion of the results
 obtained for \( \bei \, x \) are performed in Sec. 5. In Sec. 6 we
 show an immediate application to engineering. Finally, Sec. 7 aims
 its attention to discussion and conclusions of the paper.

\section {Two--point quasifractional approximation form for the real
 Kelvin function}
As it is well known, the real Kelvin function is defined as \cite{abram}

\begin{equation}
\label{ber}
\ber \, x = \Re( J_{0}(i \sqrt{i} \, x) )
\end{equation}
where $ J_{0}(z) $ is the zeroth order Bessel function.

The power series is given by

\begin{equation}
\label{ber-series}
\ber \, x = 1 - \frac{(\frac{1}{4}x^2)^2}{(2!)^2} +
 \frac{(\frac{1}{4}x^2)^4}{(4!)^2} - \cdots
\end{equation}

The asymptotic expansion of $ \ber \, x$ has been also given in
 several references. Our interest here is mainly in the leading term,
 which can be written as
 
\begin{equation}
\label{ber-asymp}
\ber \, x \sim \displaystyle \frac{e^{x/\sqrt{2}}\left(\frac{\sqrt{2+
\sqrt{2}}}{2}\cos(\frac{x}{\sqrt{2}})+\frac{\sqrt{2-\sqrt{2}}}{2}
\sin(\frac{x}{\sqrt{2}})\right)}{\sqrt{2 \pi x}}.
\end{equation}

Once the potential series (\ref{ber-series}) and the asymptotic expansion
 (\ref{ber-asymp}) have been obtained, it is possible to determine the
 form of the quasifractional approximations of the function $ \ber \, x $,
 which has an essential singularity at infinity. That kind of singularity
 is characteristic of all the hypergeometric confluent functions. The
 asymptotic expansion (\ref{ber-asymp}) picks up this singularity and
 shows it up through a branch point at infinity together with essential
 singularities at infinity of the exponential functions. Since the
 branch points come up in pairs, the leading term of the asymptotic
 expansion shows also a second branch point at $x=0$. However, the
 behaviour of the function $\ber \, x$ is regular at $x=0$, therefore
 the asymptotic form in (\ref{ber-asymp}) is not suitable as an
 approximation for $\ber \, x$ in the region near zero. In order
 to pursue the goal of picking up the right behaviour at infinity,
 but not to introduce undesired singularities in the zone of interest,
 a suitable auxiliary function has to be chosen. The auxiliary function
 should have the right ramification form at infinity, and the second
 branch point located outside the zone of interest (i.e., the negative axis)

Only powers of the form $4n$ appear in the potential series of
 $\ber \, x$. In order to achieve the adequate efficency in the
 approximation, we should find such auxiliary functions and
 fractional approximations whose potential series only have
 exponents multiple of $4$.

Since the second branch point introduced with the asymptotic
 expansion must be outside the zone of interest, it is possible
 to choose as an auxiliary function $A_{1}(x)=1/\sqrt{1+x}$;
 however, that auxiliary function would not be efficient,
 since undesired power terms as $x^2,x^3,\ldots$ would appear.
 A more suitable auxiliary function would be
 $\tilde{A}_{1}(x)=1/\sqrt[8]{1+x^4}$ or $\tilde{A}_{2}(x)=1/\sqrt[8]
{1+ \tau ^4 x^4}$. However, in the last case the parameter $ \tau $
 could be chosen in a convenient way in order to get better accuracy
 with the approximation, as we will discuss later, this can be
 considered as a free parameter.

The other singularity that must be introduced through the auxiliary
 functions is of the form $e^{x/\sqrt2}\cos \left( \frac{x}{\sqrt2}
 \right)$ and $e^{x/\sqrt2}\sin \left( \frac{x}{\sqrt2} \right)$.
 The efficiency criteria limit the auxiliary functions by forcing
 them to have exponents of the form $4n$. The above stated line
 of thought leads us to choose as convenient auxiliary functions

\begin{equation}
\label{aux_function_cos}
\cosh \left( \frac{x}{\sqrt{2}} \right) \cos
 \left( \frac{x}{\sqrt{2}} \right)
\end{equation}
and

\begin{equation}
\label{aux_function_sin}
\frac{\sinh \left( \frac{x}{\sqrt{2}} \right) \sin
 \left( \frac{x}{\sqrt{2}} \right)}{x^2}.
\end{equation}

Once the auxiliary functions are chosen as above,
 the fractional approximations automatically have
 only powers multiple of 4.

All the previous considerations lead to the following form of
 the two--point quasifractional approximation for the real
 Kelvin function 

\begin{equation}
\label{ber_aprox_form}
\tilde{\ber} \, x = \frac{\sum_{k=0}^{n} p_{k}x^{4k} \, \cosh
 \left( \frac{x}{\sqrt{2}} \right) \cos \left( \frac{x}{\sqrt{2}}
 \right) + \frac{\sqrt{1+ \alpha ^{2} x^{4}}}{x^{2}} \sum_{k=0}^{n}
 P_{k}x^{4k} \, \sinh \left( \frac{x}{\sqrt{2}} \right) \sin 
\left( \frac{x}{\sqrt{2}} \right)}{\left(1+ \sum_{k=1}^{n} q_{k}x^{4k} \right)
 \sqrt[8]{1+ \tau ^{4} x^{4}}}.
\end{equation}

The auxiliary function $ \sqrt{1+ \alpha ^{2} x^{4}} $ has to be
 introduced in order to cancel the $1/x^2$ behaviour at infinity.
 With this function we also define a second free parameter $ \alpha $.

\section {Calculation of the two--point quasifractional approximation
 to the real Kelvin function}

Here we will consider only the simplest approximation to $ \ber \, x $,
 thus $n$ will be one and the approximation in eq. (\ref{ber_aprox_form})
 will be reduced to

\begin{equation}
\label{ber_aprox}
\tilde{\ber} \, x = \frac{(p_{0}+p_{1}x^{4}) \, \cosh 
\left( \frac{x}{\sqrt{2}} \right) \cos 
\left( \frac{x}{\sqrt{2}} \right) + 
\frac{\sqrt{1+ \alpha ^{2} x^{4}}}{x^{2}} (P_{0}+P_{1}x^{4}) \, 
\sinh \left( \frac{x}{\sqrt{2}} \right) \sin \left( \frac{x}{\sqrt{2}}
 \right)}{(1+qx^4) \sqrt[8]{1+ \tau ^{4} x^{4}}}.
\end{equation}
The ideas developed on the latest papers on quasifractional
 approximations are in the sense of not to determine all the
 coefficients through the powers series and the asymptotic
 expansions but to leave one or two free parameters---$\alpha$
 and $\tau$---that can be determined by minimising the maximum
 absolute error. It is clear that other methods to measure the
 discrepancy between the approximant and the exact function
 can also be used, such as the Lebesgue's integral of
 the quadratic difference, or other more elaborated methods
 using Lebesgue's integrals of the $p$-powers of the
 difference. However, for us, the most convenient way to
 measure that discrepancy  has been the maximum absolute error.

The main reason for using free parameters is to avoid the
 characteristic defects of classic Pad{\'e} \cite{baker}, also
 common in quasifractional approximations. Defects are when
 an extraneous pole appears near a zero in the numerator.
 That is, there is a zero in the numerator in $x_{0}$ and another
 zero in the denominator in $x_{1}$ and the difference $|x_0-x_1| $
 is very small. Due to that problem, the approximation is in good
 agreement with the function in the whole zone of interest,
 except when close to $x_0$ and $x_1$, where it goes to
 $+\infty$, $-\infty$ and $0$.

In order to determine the five unknowns $p_0,P_0,p_1,P_1,q$
 three terms of the power series (\ref{ber-series}) and two
 terms of the asymptotic expansion (\ref{ber-asymp}) will be
 used. So a linear system of five equations with five unknowns
 is obtained. The solutions will be given in terms of $\alpha$
 and $\tau$.

Looking at the exponential term in (\ref{ber-asymp}), the
 variable $x/\sqrt2$ appears as more convenient. Thus the
 power series (\ref{ber-series}) will be written now as

\begin{equation}
\label{serie2}
\ber(x \sqrt2) = 1 - \frac{1}{16}x^4+\frac{1}{9216}x^8- \cdots
\end{equation}
since we are using only three terms of this series we do
 not need to go further away than $x^8$.

Now, to compare the power series of $ \ber \, x $ and $
 \tilde{\ber} \, x $, we will first multiply both functions
 by $ (1+qx^4) $ in order to rationalise the right hand side.
 Furthermore the auxiliary functions appearing in (\ref{ber_aprox})
 and defined in (\ref{aux_function_cos}) and (\ref{aux_function_sin}),
 should be replaced by their respective power series. Thus we obtain

\begin{eqnarray}
\label{system0}
(1 + q x^4)(1 - \frac{1}{16} x^4 + \frac{1}{9216} x^8) = \nonumber \\
(1 - \frac{1}{8} \tau ^{4} x^4 + \frac{9}{128} \tau ^{8} x^8) \,
 ((1 - \frac{1}{24} x^4 + \frac{1}{40320} x^8)(p_0 + p_1 x^4) +
 \nonumber \\    ( \frac{1}{2} - \frac{1}{720} x^4 +
 \frac{1}{3628800} x^8)(1 + \frac{1}{2} \alpha ^{2} x^4 - 
\frac{1}{8} \alpha ^{4} x^8) (P_0 + P_1 x^4)) + O(x^{12})
\end{eqnarray}
where the first term in the first parenthesis on the right hand
 side is the power expansion of $ \sqrt[8]{1+ \tau ^4 x^4} $, and 
constant factors have been absorbed in the definition of the parameters
$\tau$ and $q$.

Equalising now the powers $x^0$, $x^4$ and $x^8$ in both sides
 of the equation we obtain

\begin{eqnarray}
\label{system1}
p_0 + \frac{1}{2} P_0 &=& 1   \label{x0} \\
(- \frac{1}{24} - \frac{1}{8} \tau ^4) p_0 + (- \frac{1}{16} 
\tau ^4 - \frac{1}{720} + \frac{1}{4} \alpha ^2) P_0 + 
\frac{1}{2} P_1 + p_1 - q &=& - \frac{1}{16} \nonumber \\ 
\label{x4} \\
(\frac{1}{192} \tau ^4 + \frac{1}{40320} + \frac{9}{128}
 \tau ^8)p_0 + \nonumber \\ 
(\frac{1}{5760} \tau ^4 + \frac{9}{256} \tau ^8 - \frac{1}{16}
 \alpha ^4 - \frac{1}{32} \tau ^4 \alpha ^2 - \frac{1}{1440}
 \alpha ^2 + \frac{1}{3628800}) P_0 + \nonumber \\
(- \frac{1}{24} - \frac{1}{8} \tau ^4) p_1 + (-\frac{1}{16}
 \tau ^4 -\frac{1}{720} + \frac{1}{4} \alpha ^2) P_1 +
 \frac{1}{16} q &=& \frac{1}{9216} \label{x8} 
\end{eqnarray}
The leading term of the asymptotic expansion of $ \tilde{\ber} \, x $ is

\begin{equation}
\label{ber_aprox_asymp}
\tilde{\ber} \, x \sim e^{x / \sqrt{2}} \left[ \frac{1}{2}
 \frac{p_1 \cos \left( \frac{x}{ \sqrt2} \right) + P_1 \alpha
 \sin \left( \frac{x}{ \sqrt2} \right)}{q \sqrt{ \tau x}} \right].
\end{equation}
And comparing now the leading terms of $ \ber \, x $, we obtain

\begin{eqnarray}
\label{system2}
p_1 &=& \frac{ \sqrt{2+ \sqrt2} \sqrt{ \tau } q}{ \sqrt{2 \pi }}
  \label{inf1} \\
P_1 &=& \frac{ \sqrt{2- \sqrt2} \sqrt{ \tau } q}{ \sqrt{2 \pi }\alpha }
 \label{inf2}
\end{eqnarray}

Equations (\ref{x0}), (\ref{x4}), (\ref{x8}), (\ref{inf1}) and
 (\ref{inf2}) determine the values of $p_0,P_0,p_1,P_1$ and $q$
 as functions of $ \alpha $ and $ \tau $. 

When these equations are solved, an expression for each parameter
 is obtained in terms of $ \alpha $ and $ \tau $. And an
 approximation $\tilde{\ber}(x, \alpha , \tau)$ is defined
 for each value of $ \alpha $ and $ \tau $.

Given the initial values $\alpha _0$ and $\tau _0$, we
 numerically determine the value of the function $|\tilde{\ber}
(x, \alpha , \tau) - \ber(x)|$ and we select the maximum value 
 of this function, which will be the maximum absolute error.
 Therefore, the maximum absolute error, $\epsilon ( \alpha , \tau )$,
 will be a function of $ \alpha $ and $ \tau $. Looking now at $\epsilon
 ( \alpha , \tau )$ as a function, we can determine the values
 $\alpha _m$ and $\tau _m$ which minimise this function as a
 two variable function. That procedure has been followed and
 the values $\alpha _m$ and $\tau _m$ are found as

\begin{eqnarray}
\label{value1}
\alpha_m &=& 0.98 \nonumber \\
\tau_m &=& 0.8367 \nonumber
\end{eqnarray}

In order to avoid the defect in the approximation we have to consider
 only positive values of $q$, thus there is no problem with positive
 values for the variable $x$ which is the region of interest.
 This must be taken into account when sweeping through $\alpha$ and
 $\tau$ in the minimising procedure. Local minima which yield negative
 values for $q$ must be discarded.

Using these values of $ \alpha $ and $ \tau $ in (\ref{x0}), (\ref{x4}),
 (\ref{x8}), (\ref{inf1}) and (\ref{inf2}), the parameters
 $p_0,P_0,p_1,P_1$ and $q$ are determined and the results are

\begin{eqnarray}
\label{value2}
q &=& 27627.311660 \nonumber \\
p_0 &=&  -9750.649914 \nonumber \\
P_0 &=&  19503.300340 \nonumber \\
p_1 &=& 18628.544300 \nonumber \\
P_1 &=& 7873.669071 \nonumber 
\end{eqnarray}

\section{Accuracy of the two--point quasifractional approximation
 of the real Kelvin function}

As it has been stated before, the discrepancy of the approximation will
 be determined by $ \Delta \ber \, x = \ber \, x - \tilde{\ber} \, x $. 

A plot of the real Kelvin function, together with the approximation
 and ten times the absolute error is displayed in Figure~1.

Besides, a table of the first five roots of the Kelvin function and
 the approximation is introduced.

Notice that the highest relative error in the roots is the first one.
 Quasifractional approximations to any function have normally their
 worst accuracy for values of order one. The behaviour of the
 approximation becomes more accurate as $x$ increases.

Another two measures of the good agreement of the approximation
 with the real Kelvin function is the relative difference of the
 position in $x$ of the maxima of the function and the relative
 error of the amplitude of each maximum. The following table shows
 the position in $x$ of the first five maxima of $\ber \, x$ and
 $\tilde{ \ber } \, x$ as $\ber ' \, x$ and $\tilde{ \ber } ' \,
 x$ respectively, and the relative difference between them. The
 value of each of the first five maxima is also shown, appearing
 as $\ber  _m \, x$ and $\tilde{ \ber } _m \, x$ for the real
 Kelvin function and the approximation respectively. The relative
 error of those figures also appears.

Notice that all the relative errors, no matter what is being
 measured (i.e., the roots, the position of the maxima, the
 amplitude) decrease as $x$ becomes greater.

\section{Calculation of the two--point quasifractional
 approximation for the Imaginary Kelvin function}
The same line of thought carried out for the real Kelvin function
 will be followed in this section, since the form of the asymptotic
 expansions and the power series for both $\ber \, x$ and $\bei \,
 x$ are very similar.

The imaginary Kelvin function is defined as \cite{abram}

\begin{equation}
\label{bei}
\ber \, x = \Im( J_{0}(i \sqrt{i} \, x) )
\end{equation}
where $ J_{0}(z) $ is the zeroth order Bessel function.
The power series of $\bei \, x$ is given by

\begin{equation}
\label{bei-series}
\bei \, x = \frac{1}{4} x^2 - \frac{(\frac{1}{4}x^2)^3}{(3!)^2}
 + \frac{(\frac{1}{4}x^2)^5}{(5!)^2} - \cdots
\end{equation}
The asymptotic expansion of $ \bei \, x$ has been also given in
 several references. Again, our interest here lays mainly on the
 leading term, which can be written as
 
\begin{equation}
\label{bei-asymp}
\bei \, x \sim \displaystyle \frac{e^{x/\sqrt{2}}
\left(\frac{\sqrt{2+\sqrt{2}}}{2}\sin(\frac{x}{\sqrt{2}})
-\frac{\sqrt{2-\sqrt{2}}}{2}\cos(\frac{x}{\sqrt{2}})\right)}
{\sqrt{2 \pi x}}.
\end{equation}

It is a characteristic of hypergeometric confluent functions
 to have an essential singularity at infinity. The asymptotic
 expansion picks that singularity up and shows it through a
 branch point at infinity with the exponential functions, in
 the same way that happened for $\ber \, x$.

The power series of $\bei \, x$ goes as $x^2,x^6,x^{10}, 
\ldots $. In order to have an efficient approximation, the
 auxiliary functions chosen for moving the artificially
 introduced branch point out of the positive axis and for
 introducing the singularities that appear in the asymptotic
 expansion should have a power series of the same form as $\bei \, x$.

Following the same stream of thought used in determining the
 form of $\ber \, x$, a correct form for the quasifractional
 approximation of $\bei \, x$ is

\begin{equation}
\label{bei_aprox_form}
\tilde{\bei} \, x = \frac{\frac{x^{2}}{\sqrt{1+ \bar{ \alpha }
 ^{2} x^{4}}} \sum_{k=0}^{n} \bar{p}_{k}x^{4k} \, \cosh
 \left( \frac{x}{\sqrt{2}} \right) \cos \left( \frac{x}{\sqrt{2}}
 \right) +  \sum_{k=0}^{n} \bar{P}_{k}x^{4k} \, \sinh \left(
 \frac{x}{\sqrt{2}} \right) \sin \left( \frac{x}{\sqrt{2}}
 \right)}{\left(1+ \sum_{k=1}^{n} \bar{q}_{k}x^{4k} \right)
 \sqrt[8]{1+ \bar{ \tau } ^{4} x^{4}}}
\end{equation}
where $\bar{ \alpha }$ and $\bar{ \tau }$ are free parameters,
 introduced for the same reason explained for $\ber \, x$.
We are going to use only the simplest form of the approximation,
 thus $n=1$ will be substituted in the previous equation.
 Thus the two--point quasifractional approximation of $\bei
 \, x$ will give

\begin{equation}
\label{bei_aprox}
\tilde{\bei} \, x = \frac{\frac{x^{2}}{\sqrt{1+ \bar{ \alpha }^{2}
 x^{4}}}( \bar{p}_{0}+ \bar{p}_{1}x^{4}) \, \cosh \left(
 \frac{x}{\sqrt{2}} \right) \cos \left( \frac{x}{\sqrt{2}} \right)
 +  ( \bar{P}_{0}+ \bar{P}_{1}x^{4}) \, \sinh \left( \frac{x}
{\sqrt{2}} \right) \sin \left( \frac{x}{\sqrt{2}} \right)}
{(1+ \bar{q} x^4) \sqrt[8]{1+ \bar{\tau} ^{4} x^{4}}}.
\end{equation}

From now on, exactly the same procedure has been followed in
 determining the set of five equations used for finding the
 values of the parameters $ \bar{p}_0, \bar{P}_0, \bar{p}_1,
 \bar{P}_1, \bar{q}$ in terms of $\bar{ \alpha }$ and
 $\bar{ \tau }$, and for minimising the discrepancy of the approximation.

When dealing with quasifractional approximations it is common
 to take as an initial ansatz $\alpha$ and $\tau$ equal to 1.
 This choice proved to be a valid starting point for $\tilde{\ber}
 \, x$, even though smaller differences are obtained when using the
 minimised $\bar{ \alpha_m }$ and $\bar{ \tau_m }$ given in the
 previous section. The $\tilde{\bei} \, x$ function is somewhat
 more pathological since the choice $\bar{ \alpha } =
 \bar{ \tau } = 1$ yields a negative value of $\bar{q}$,
 making the initial ansatz useless.

For minimising the maximum absolute error of the two--point quasifractional
 approximation of $\bei \, x$ a downhill simplex method has
 been used \cite{recipes}. The values obtained for the parameters are

\begin{eqnarray*}
\label{value3}
\bar{ \alpha_m } &=& 3.00 \\
\bar{ \tau_m } &=& 3.00  \\
\bar{q} &=& 19.11054940  \\
\bar{p}_0 &=&  -7.21235948  \\
\bar{P}_0 &=&  15.42471896  \\
\bar{p}_1 &=& -30.32038957  \\
\bar{P}_1 &=& 24.39996523 
\end{eqnarray*}

The plots of $\bei \, x, \tilde{ \bei } \, x$ and ten times the
 absolute error ($10 \Delta \tilde{ \bei } \, x=10( \tilde{ \bei }
 \, x - \bei \, x)$ ) are shown in Figure~2.

The relative error of the first root is smaller than the one of the
 first root of $\ber \, x$ because the value of the variable $x$ is
 greater (i.e., the first root of $\bei \, x$ it is not as close to
 zero as the one of $\ber \, x$). The accuracy for all the roots of 
$\ber \, x$ and $\bei \, x$ is very high in any case.

Once again, the relative error related to the position of the first
 five maxima of the function and the error of their amplitude will
 be calculated in the same way performed for $\ber \, x$. The 
notation is exactly the same followed in the calculation of error
 for $\ber \, x$ before in Sec. 4, but adapted for the function
 $\bei \, x$.

\section{Analytic approximation to the current distribution within
 a cylindrical conductor}

The expression of the current distribution within a cylindrical 
conductor due to a travelling wave has been given in many references.
 The current distribution depends on the Kelvin real and imaginary
 functions and is given by \cite{marion}

\begin{equation}
\label{current}
\left| \frac{J_z(r)}{J_s} \right| = \left( \frac{ \ber ^2
 ( \sqrt2 r/ \delta) + \bei ^2 ( \sqrt2 r/ \delta) }{ \ber ^2
 ( \sqrt2 r_0/ \delta) + \bei ^2 ( \sqrt2 r_0/ \delta) } \right) ^{1/2}
\end{equation}
where $J_s$ is the current density on the surface of the conductor and
$r_0$ is the radius on the surface. The skin depth is given by
$\delta$, and it depends on the frequency of the propagating wave.

By means of substitution of $\tilde{ \ber } \, x$ and $\tilde{ \bei }
\, x$ in (\ref{current}) it is possible to get an analytical
approximation to the current distribution within a cylindrical
conductor.

The analytical approximation obtained has been used for studying the
current distribution within a copper wire with $r_0=0.5$mm. Four
different values of the frequency have been taken. The same figures
and plots could be found in Marion, for comparison purposes.

The plots obtained using our approximation and those obtained using
numerical computation are coincident and no difference can be found at
this scale. Therefore, we are only showing the errors in Figure~3.

\section{Conclusions}
In this paper analytic approximations have been found for the function
$\ber \, x$ and $\bei \, x$. These have been used to calculate the
current distribution within a cylindrical conductor produced by the
propagation of one wave of frequency $\omega $.

In spite of the strong fluctuation characteristic of the functions
$\ber \, x$ and $\bei \, x$, the simple approximations found here not
only reproduce the function with high accuracy for all positive values
of the variable but the infinite roots or zeroes of the function are
also obtained and the relative error of each root decreases with the
magnitude of the root. The largest relative error of the roots for
$\ber \, x$ is about 2\% and for $\bei \, x$ is 0.5\%.

The measure of the discrepancy is difficult for these functions because the
function fluctuates and, furthermore, the maximum of the amplitude
increases and goes to infinity with $x$. Therefore, if we consider the
absolute error this will go to infinity with $x$.

On the other hand, if we study the relative error, these errors will
go to infinity at the roots of the function. The best way to measure
the good agreement of our approximation is just through the errors of
the zeroes and of the amplitude of the oscillations. It is significant
that these errors decrease with $x$ despite of the large value of the
function. The maximum error of the amplitude for $\ber \, x$ is 3\%,
and 4\% for $\bei \, x$ in the value of the function and of 0.28\% and
0.80\% in the site of the maximum for $\ber \, x$ and $\bei \, x$
respectively.
 
These errors clearly show that our approximation can be used for most
of the applications we know until now, for example thermo--electrical
corrosion in cylindrical conductors. In general, these approximations
might be used whenever Kelvin functions appear and a numerical 
solution is not appropriate.  As an example, we have applied our method
to the simplest case of the current distribution in a
cable, obtaining quite accurate results.

\newpage

\begin{table}[ht]
\caption{Comparison between the first five roots of $ \ber \, x $
 and $ \tilde{\ber} \, x $ }
\begin{tabular}{lcr}

$ \ber \, x $ roots & $ \tilde{\ber} \, x $ roots & Relative error
 ( \% ) \\  \hline
2.84892    & 2.78620    & 2.2    \\  
7.23883    & 7.22030    & 0.26    \\  
11.67396    & 11.66266    & 0.094    \\  
16.11356    & 16.10548    & 0.052    \\  
20.55463    & 20.54834    & 0.032    \\ 
\end{tabular}
\end{table}

\begin{table}[ht]
\caption{Comparison between the first five maxima of $ \ber \, x $ and
 $ \tilde{\ber} \, x $ and their derivatives.}
\begin{tabular}{cccccc}

$ \ber ' \, x $ & $ \tilde{\ber} ' \, x $ & err.(\%) & $\ber  _m \, x$ &
 $\tilde{ \ber } _m \, x$ & err.(\%)\cr  \hline
6.03871   & 6.0215   & 0.28  & -8.86404  & -8.61484 & 2.9\cr  
10.51364  & 10.5027  & 0.10  & 153.782   & 151.237  & 1.7\cr  
14.96844  & 14.9605  & 0.053 & -2968.68  & -2933.93 & 1.2\cr  
19.41758  & 19.4114  & 0.032 & 60161.2   & 59616.8  & 0.92\cr  
23.86430  & 23.8592  & 0.021 & $-1.25374 \times 10^6$ & $-1.2445
 \times 10^6$ & 0.75 \cr 
\end{tabular}
\end{table}

\begin{table}[ht]
\caption{Comparison between the first five roots of $ \bei \, x $ and
 $ \tilde{\bei} \, x $. }
\begin{tabular}{lcr}

$ \bei \, x $ roots & $ \tilde{\bei} \, x $ roots & Relative error ( \% ) \\ 
 \hline
5.02622    & 4.99873    & 0.55    \\  
9.45541    & 9.44110    & 0.15    \\  
13.89349    & 13.88400    & 0.069    \\  
18.33398    & 18.32689    & 0.037    \\  
22.77544    & 22.76977    & 0.024    \\ 
\end{tabular}
\end{table}

\begin{table}[ht]
\caption{Comparison between the first five maxima of $ \bei \, x $ and $
 \tilde{\bei} \, x $ and their derivatives.}
\begin{tabular}{cccccc}

$ \bei ' \, x $ & $ \tilde{\bei} ' \, x $ & err.(\%) 
& $\bei  _m \, x$ & $\tilde{ \bei } _m \, x$ & err.(\%)\\  \hline
3.77320   & 3.74307   & 0.80  & 2.34615  & 2.259    & 3.9 \\  
8.28099   & 8.26718   & 0.17  & -36.1654 & -35.4089 & 2.2 \\  
12.74215  & 12.7329   & 0.073 & 670.16   & 660.955  & 1.4 \\  
17.19343  & 17.1865   & 0.040 & -13305.5 & -13169.6 & 1.1 \\  
21.64114  & 21.6356   & 0.026 & 273888   & 271661   & 0.82  \\ 
\end{tabular}
\end{table}

\begin{table}[ht]
\caption{Skin depth at different frequencies. }
\begin{tabular}{lcr}

Case & $ \nu = \omega / 2 \pi $ & $\delta$ (mm) \\  \hline
1    & $10^3$   & 2.1    \\  
2    & $10^4$   & 0.66    \\  
3    & $10^5$   & 0.21    \\  
4    & $10^6$   & 0.066    \\ 
\end{tabular}
\end{table}

\newpage
\begin{figure}

\centerline{\includegraphics[width=2.7in]{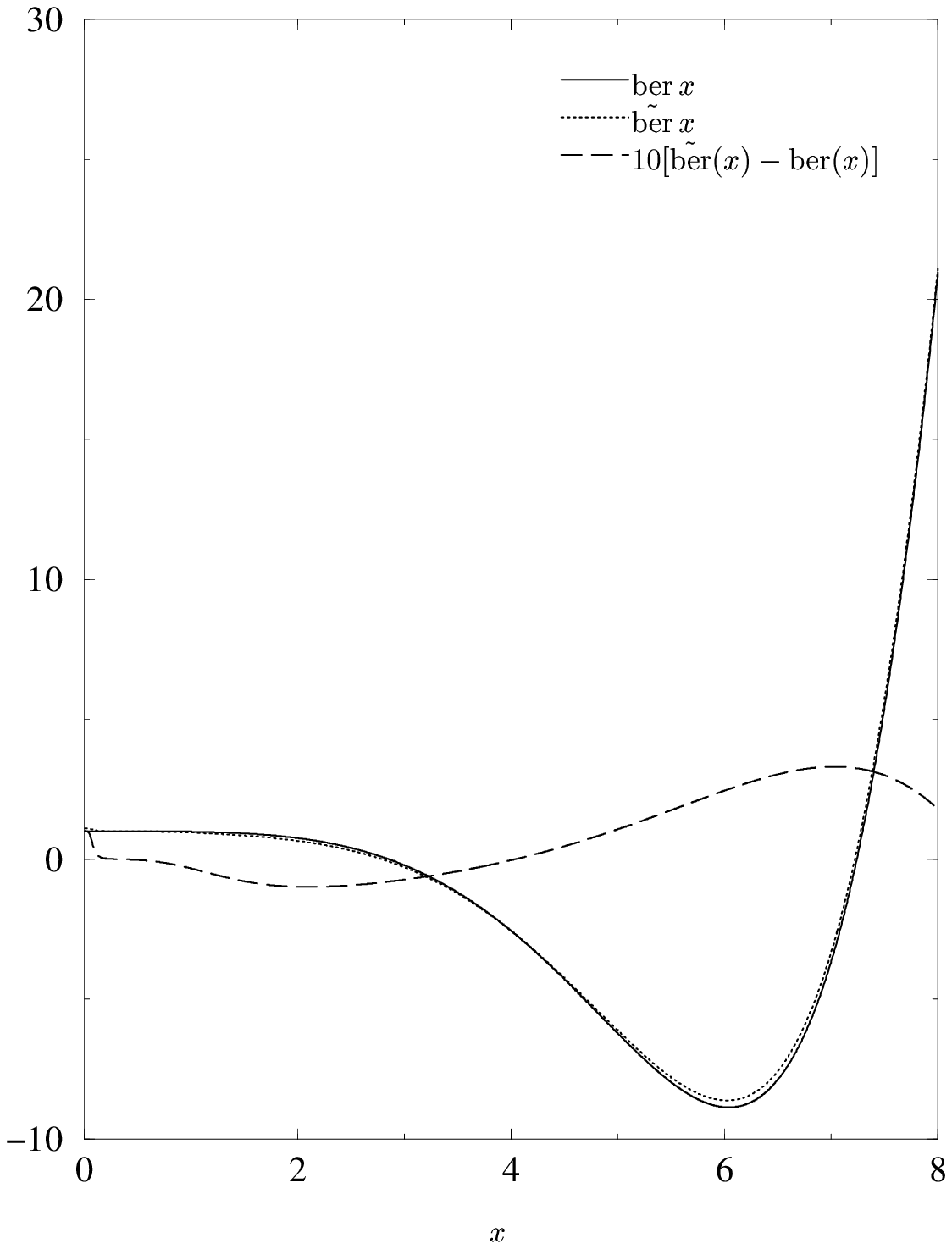}}
\caption{Kelvin real function and its two--point quasifractional approximation, along with ten times the discrepancy between them.}
\label{fig: berx}

\end{figure}
\begin{figure}

\centerline{\includegraphics[width=2.7in]{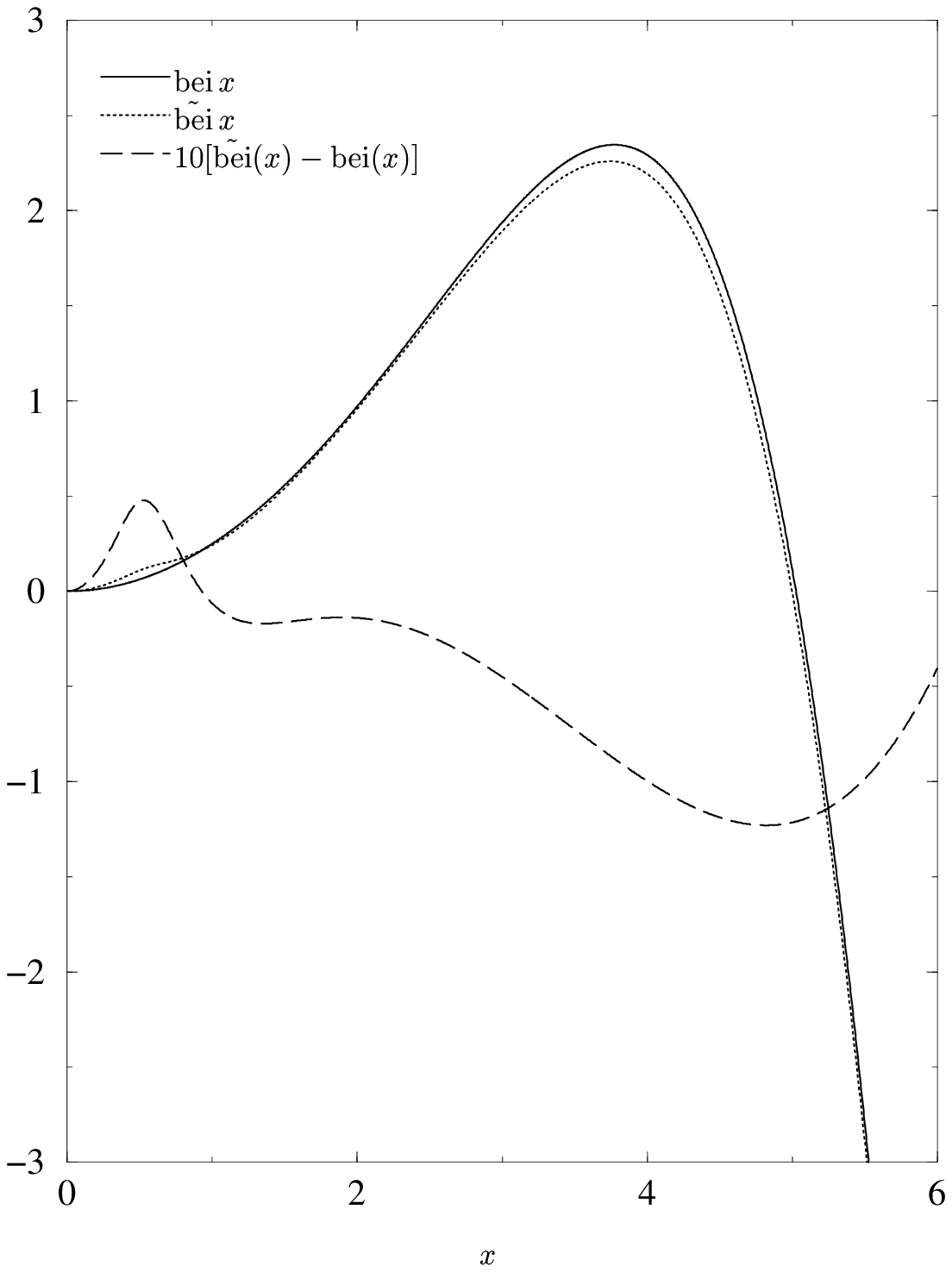}}
\caption{Kelvin imaginary function and its two--point quasifractional approximation, along with ten times the discrepancy between them.}
\label{fig: beix}

\end{figure} 
\begin{figure}

\centerline{\includegraphics[width=2.7in]{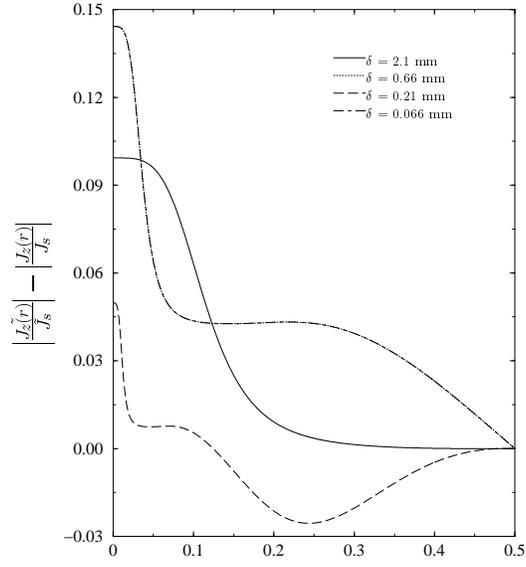}}
\caption{Plots of the discrepancy between the analytical current 
distribution within a cylindrical
 conductor and the results obtained with two--point quasifractional
approximation for four different values of the skin depth.}
\end{figure} 
\end{document}